\documentclass[%
 reprint,
 amsmath,amssymb,
prb,
nolongbibliography
]{revtex4-2}

\usepackage{graphicx}
\usepackage{color}
\usepackage{graphicx,subfigure}
\usepackage{dcolumn}
\usepackage{bm}
\usepackage[colorlinks,citecolor=blue,urlcolor=blue,linkcolor=blue,bookmarks=false,hypertexnames=true]{hyperref} 

\newcommand{\ie}{{\sl i.e.}}

\newcommand{\dztrt}{$d_{3z^2-r^2}$}
\newcommand{\dxtyt}{$d_{x^2-y^2}$}
\newcommand{\dxy}{$d_{xy}$}
\newcommand{\dxz}{$d_{xz}$}
\newcommand{\dyz}{$d_{yz}$}
\newcommand{\femgo}{Fe$_{1}$/(MgO)$_{3}$(001)}
\newcommand{\mub}{$\mu_{\rm B}$}
\newcommand{\ef}{$E_{\rm F}$}

\begin{document}

\preprint{APS/123-QED}

\title{Anisotropic carrier dynamics in a laser-excited Fe$_{1}$/(MgO)$_{3}$(001) heterostructure from real-time time-dependent DFT}

\author{Elaheh Shomali, Markus E. Gruner and Rossitza Pentcheva}

\affiliation{Department of Physics and Center for Nanointegration Duisburg-Essen (CENIDE), University of Duisburg-Essen, Lotharstr. 1, 47057 Duisburg, Germany}
\date{\today}

\begin{abstract}

The interaction of a femtosecond optical pulse with a Fe$_{1}$/(MgO)$_{3}$(001)
metal/oxide  heterostructure is investigated using time-dependent density functional theory (TDDFT) calculations in the real-time domain.
We systematically study electronic excitations
as a function of laser frequency, peak power density and polarization direction.
While spin-orbit coupling is found 
to result  in only a small time-dependent reduction of magnetization (less than 10\,\%), we find a marked 
anisotropy in the response to in-plane and out-of-plane polarized light, which changes its character
qualitatively depending on the excitation energy:
the Fe-layer is efficiently addressed at low frequencies by in-plane polarized light,
whereas for frequencies higher than the MgO band gap, we find a particularly strong response of the central MgO-layer for cross-plane polarized light.
For laser excitations between the charge transfer gap and the MgO band gap, the interface plays the most important
role, as it mediates concerted transitions from the valence band of MgO into the $3d$ states of Fe closely above the Fermi level
and from the Fe-states below the Fermi level into the conduction band of MgO.
As these transitions can occur simultaneously altering charge balance of the layers,
they could potentially lead to an efficient transfer of excited carriers into the MgO bulk, 
where the corresponding electron and hole states can be separated by
an energy which is significantly larger than the photon energy.
\end{abstract}
\maketitle


\section{\label{sec:level1}Introduction}
The recent development in brilliant ultrafast optical and x-ray sources with femtosecond resolution
has fostered investigations that aim at fundamental understanding of light-matter interaction and the resulting non-equilibrium properties of matter \cite{cn:Cavalleri05,cn:Ernstorfer09,cn:Melnikov11,cn:Ichikawa11,cn:Stojchevska14,cn:Huebener17,cn:Lloyd21}. This paves the way for applications such as light-induced control of magnetization reversal for data storage  \cite{cn:Stamm07,cn:Radu11} or
light-induced hot carriers for photo-catalytic processes, detection devices and energy
harvesting \cite{cn:Brongersma15,cn:Narang16}.

Time-dependent density functional theory (TDDFT) \cite{cn:Marques12,cn:Sharma14,cn:Olevano18} in the real-time domain has evolved
as an important tool to unravel the dynamics of excitation processes
in organic and inorganic materials \cite{cn:Lloyd21}.
Recent efforts cover various aspects of
light-matter interaction in metals \cite{cn:Volkov19,cn:Senanayake19} and semiconductors \cite{cn:Sato14b,cn:Tancogne17,cn:Pemmaraju18},
the simulation of pump-probe experiments in Si \cite{cn:Sato14a} or
nonlinear absorption in complex molecules \cite{cn:Guandalini21} as well as
charge and energy transfer after the impact of fast ions in aluminum sheets and nanoclusters \cite{cn:Schluenzen19,cn:Deuchler20,cn:Kononov20}.

Many potential applications of non-equilibrium states induced by laser pulses involve a combination of different materials, for instance in the field of photo-catalysis \cite{cn:Wu12,cn:Akimov13} or
optically induced (de-)magnetization processes and spin-transport
in multilayer materials for magnetic storage \cite{cn:Rudolf12,cn:Cardin20}.
In this spirit not only experiment but also first-principles studies on optically excited systems
in the real-time domain are shifting from single material and bulk systems towards heterogeneous systems with increasing complexity.
While significant fraction of this research is concerned with the fundamental understanding of 
demagnetization processes in metallic multilayer systems \cite{dewhurst2018laser,cn:Chen19},
the photo-induced catalytic processes and energy conversion usually involve
the transfer of energy and charge across interfaces \cite{cn:Akimov13,cn:Wu15}.
Here, real-time TDDFT (RT-TDDFT) approaches
contribute towards a fundamental understanding of the carrier dynamics related to
the plasmon-mediated injection of hot electrons from metallic nanostructures into semiconductors or
insulators \cite{cn:Long14,cn:Zhang19,cn:Iida20,cn:Tomko21}.
The efficient transfer of excitations or dissipation of energy across interfaces
has also evolved as an important topic of ultra-fast pump-probe experiments
\cite{cn:Melnikov11,cn:Rothenbach19,cn:Schumacher19,beyazit2019local}.

Understanding the interaction of laser pulses with a  heterostructure offers the possibility
to induce selectively excitations in a particular layer which may propagate into the entire system.
At the interface between metallic and insulating layers, the electronic structure is characterized by the hybridization between orbitals of the metallic and insulating part, which usually
leads to localized states within the gap in the adjacent layer of the insulator \cite{cn:Oleinik00,cn:Butler01,cn:Abedi10,cn:Rothenbach19}.
As we will show, these interface states might be employed by a proper choice of the photon energy
to foster a simultaneous, concerted excitation of electrons and holes. This allows for a
charge-neutral, but asymmetric propagation of the excited carriers into the metallic and insulating subsystems.
The symmetry breaking at the interface leads to splitting of states with in-plane and out-of-plane orbital character.
As pointed out earlier, changing the polarization of the incident light wave may result in a substantially different response \cite{gruner2019dynamics}, which could be used to select particular excitations at the interface and
differentiate between the pathways of energy propagation into the bulk materials.

Fe/MgO(001) represents the ideal model system to explore such effects.
The Fe/MgO system was extensively investigated in the context of TMR (tunnel magnetoresistance) and benefits from the fact that electronic, magnetic and transport properties are well established  \cite{cn:Butler01,cn:Mathon01,cn:Tiusan04,cn:Waldron06,cn:Belashchenko05,cn:Heiliger08,cn:Peralta08,cn:Rungger09,PhysRevB.79.174414,cn:Abedi10}. Recently, also optical and lattice excitations have been subject to
theoretical and experimental studies \cite{cn:Rothenbach19,gruner2019dynamics,cn:Eggert20,cn:rothenbach2021effect}.
Bulk Fe is a ferromagnetic metal with bcc structure
and a magnetic moment of $2.22\,$\mub/atom \cite{cn:LandoltBornstein2014},
showing a substantial density of occupied and unoccupied $d$-states in the
vicinity of the Fermi level.
MgO in turn is a wide band gap insulator, with an experimental band gap of
of about $\sim$$\,7.7\,$eV \cite{roessler1967electronic,whited1973exciton}).
DFT calculations with local or semi-local exchange-correlation functionals yield
$4.5\,$eV$-4.9\,$eV \cite{cn:wang2004electronic,cn:schleife2006first,cn:shishkin2007self,cn:nourozi2019electronic}
but the band gap can be improved by using hybrid exchange-correlation functionals,
for a correct description
considering quasiparticle and excitonic effects
is essential \cite{cn:shishkin07,cn:Schleife09,begum2021theoretical}.
On the other hand, since
quasiparticle corrections essentially lead to a rigid shift
of the electronic states of the insulator \cite{begum2021theoretical} in MgO,
semi-local functionals still allow for a qualitative analysis of
optical excitation processes.

\begin{figure}
\includegraphics[width=\columnwidth]{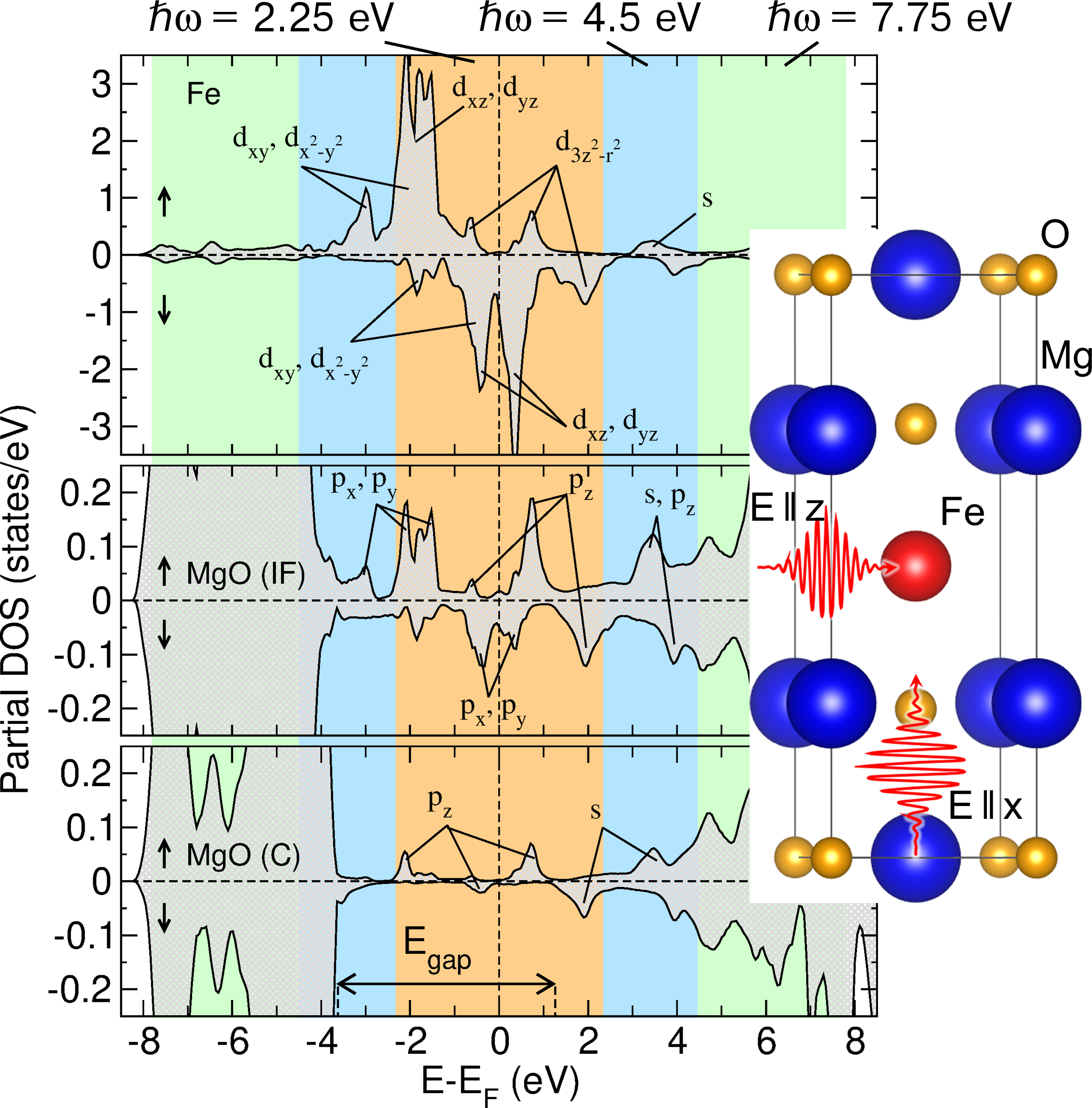}%
\caption{\label{fig:intro}
  Layer resolved electronic density of states indicating the main orbital character of the states in the vicinity of the gap. The colored areas (orange for $\hbar\omega=2.25\,$eV, blue for $\hbar\omega=4.5\,$eV, green for $\hbar\omega=7.75\,$eV) mark the maximum energy range, in which direct resonant excitations can be expected  (i.\,e.\ from the occupied states at $-\hbar\omega$ to the Fermi level and from the Fermi-level up to the unoccupied states at $\hbar\omega$). The horizontal arrow in the lowest panel denotes the position of the MgO band gap, obtained from the comparison of the bulk LDA band-structure
with the bands originating from the center layer of MgO \protect\cite{gruner2019dynamics}.
Inset: Side view of the minimal heterostructure \femgo{}. The oscillating arrows illustrate the two orientations of the propagation direction of the applied laser pulses and the corresponding polarization of the electric field with respect to the layer stacking.
}
\end{figure}

Here, we investigate a minimum four-layer system \femgo, see Fig.~\ref{fig:intro},
which enables us
of obtaining
a qualitative picture of the various types of excitations
and the dynamics of excited carriers at the metal-insulator interface in the real-time domain.
Despite its simplicity our minimum model system 
 already offers the necessary ingredients, such as a metallic and a nearly insulating layer
separated by an interface layer characterized by a considerable hybridization
between the $3d$-states of Fe and the $2p$-states of the apical oxygen, as illustrated
in the layer- and orbital-projected density of states (DOS) in
Fig.~\ref{fig:intro}(b).
We use a RT-TDDFT modelling employing the adiabatic LDA (ALDA) for exchange and correlation for
an adequate compromise regarding computational efficiency and numerical stability.
Since our investigation is limited to the first $50$\,fs and photon
energies far above the highest phonon modes, we neglect at the current stage ionic
motion and electronic dissipation channels not covered by the ALDA.

In our previous  investigation \cite{gruner2019dynamics}, we concentrated on optical excitations
with laser frequencies up to 3.27\,eV. These are still below the calculated bulk band gap of MgO
obtained from (semi-)local DFT calculations.
These photon energies allow
thus for excitations in the Fe-layer that reach beyond the charge transfer gap but not for direct excitation in MgO, see Fig.~\ref{fig:intro}.
In the present work, we now focus on laser frequencies in the vicinity and above the
LDA band gap of MgO.
We demonstrate that the excitation of the system (Fe vs.\ MgO vs.\ interface)
depends strongly on the excitation energy and the polarization direction of the electric field.
While the propagation of excited carriers across the interface could lead to
to the accumulation or depletion of charge in particular zones, 
we observe, that interface-mediated simultaneous (but independent) excitation processes,
involving electrons and holes,
can effectively compensate the net electrical transfer.
In addition, we study the effect of field strength
on the transfer of excitations and the impact of spin-orbit
interaction (SOI), which results in
transient changes in magnetization, arising from the optical excitation.

The paper is structured as follows: In Sec.\ \ref{sec:comp} we present the computational
methodology and details. Sec.\ \ref{sec:mdyn} is devoted to magnetization dynamics,
while in Sec.\ \ref{sec:inplane} we compare the excitation patterns for
three frequencies and in-plane polarization of the light field. Sec.\ \ref{sec:outplane} addresses the
dependence on the polarization direction of the electric field. Finally,
Sec.\ \ref{sec:sum} provides a discussion and summary of the results.

\section{\label{sec:comp} Computational details}

The electronic structure and time-dependent properties were obtained from density functional theory (DFT) calculations using the ELK code \cite{ELK}, which
is an all-electron full-potential linearized augmented-plane wave (LAPW) code that implements time-dependent DFT (TDDFT) in the real-time (RT) domain.
To model \femgo\ we used muffin tin radii of 1.139 \AA, 1.164 \AA\ and 0.855 \AA\ for Fe, Mg and O, respectively. The plane wave cut-off parameter, $RK_{max}$, was
set to 7. A $8 \times 8 \times 3$ $k$-mesh was used for the reciprocal space sampling.
In the ground-state calculations, the convergence criterion for the
electronic self-consistency cycle was a root-mean-square change of 10$^{-7}\,$a.u.\ in the Kohn-Sham potential.
For the exchange-correlation
functional we have chosen the local (spin) density approximation (LDA) in the parametrization of Perdew and Wang (PW92)
\cite{perdewB45}. The optimized geometry \cite{gruner2019dynamics} was previously obtained with the VASP code \cite{cn:VASP1,cn:VASP2} using the generalized gradient
approximation (GGA) of Perdew, Burke, and Ernzerhof (PBE) 
\cite{cn:Perdew96}. 
VASP (PBE) and Elk (LDA) lead to similar spin- and layer-resolved ground-state electronic partial DOS
(PDOS) of the \femgo\ heterostructure, which is shown in Fig.~\ref{fig:intro}
(for more details see~\cite{gruner2019dynamics}).

In our investigation, we simulate laser pulses with different laser frequencies and peak power densities but constant duration. The monochromatic electromagnetic wave is folded with a Gaussian envelope with a constant full-width at half-maximum (FWHM) of $5.81\,$fs. The peak of the pulse is reached at $t=11.6\,$fs after the start of the simulation.
\\
The real-time TDDFT method propagates the electron density in time by integrating the time-dependent Kohn-Sham equation (TDKS) \cite{krieger2015laser, elliott2016ultrafast, dewhurst2018laser, elliott2016optimal, dewhurst2016efficient}. The electric field of the laser pulse, expressed by the vector potential, $\bold A_{\rm ext}(t)$, enters the KS equation as a velocity gauge.
\\
By solving the TDKS equations, we can obtain the time-dependent electronic properties of a system such as time-resolved DOS (TDDOS), using the following equation, see Ref. \cite{dewhurst2018laser}:

\begin{equation}
\begin{aligned}
D_{\sigma}(E,t) = \sum_{i=1}^{\infty} \int_{\rm BZ} d^3\,\bold k \delta(E - \varepsilon_{i\bold{k}\sigma})\, g_{i\bold{k} \sigma}(t) \label{eq:TDDOS}
\end{aligned}
\end{equation}

Where $g_{i\bold{k}\sigma}(t)$ are the time-dependent and spin-resolved occupation numbers, defined as:

\begin{equation}
\begin{aligned}
g_{i\bold{k}\sigma}(t) = \sum_{j} n_{j\bold{k}\sigma} \left | \int d^3r\, \Phi_{j\bold{k}\sigma}(\bold r,t)\, \Phi^*_{i\bold{k}\sigma}(\bold r,0) \right |^2
\end{aligned}
\end{equation}

Here $n_{j\bold{k}\sigma}$ is the occupation number of the $j^{th}$ orbital and $\Phi_i$ are the
ground-state Kohn-Sham orbitals~\cite{elliott2016ultrafast}.

\begin{figure}[!htp]
\includegraphics[width=0.5\textwidth]{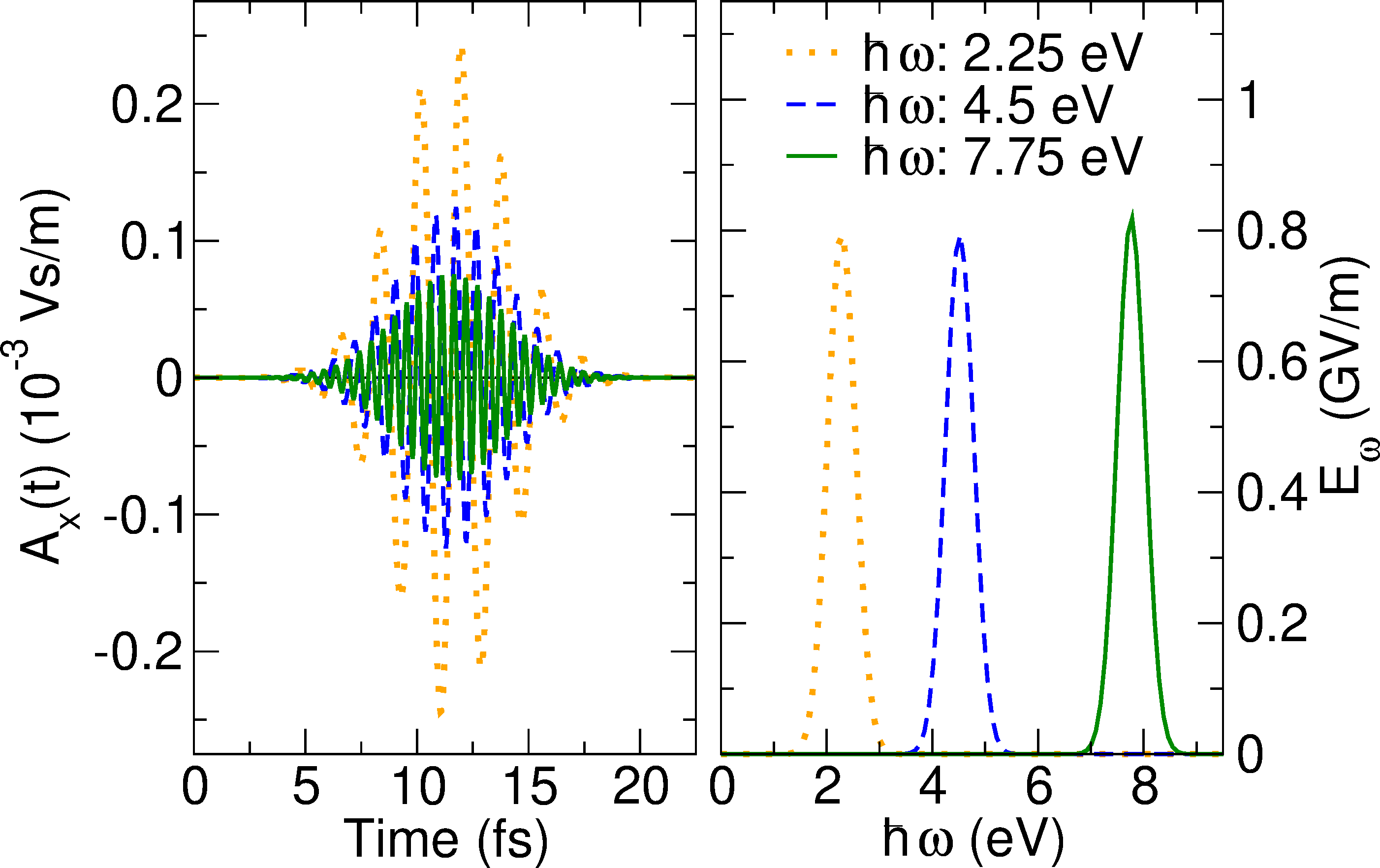}%
\caption{
  Time-dependent vector potentials of laser pulses with
different frequencies of $\hbar\omega$ = 2.25 eV, $\hbar\omega$ = 4.5 eV and
$\hbar\omega$ = 7.75 eV (left) and Fourier transform of the time-dependent electric
field for these laser pulses (right).}
\label{E-t-fourier}
\end{figure}
\section{\label{sec:Results} Results}

We carried out a systematic investigation of the impact of optical pulses with varying frequency, intensity and polarization on a \femgo\ heterostructure which consists of one
layer of Fe and three layers of MgO: two interface (IF) and one central (C) in the latter, as illustrated in Fig.~\ref{fig:intro}.
Consistent with
our previous study \cite{gruner2019dynamics}, the laser pulses were designed to achieve a fluency
typically found in experiments \cite{cn:Bierbrauer17,cn:Eschenlohr17} with
a constant peak power density $S_{\rm peak}\approx 5\times 10^{12}\,$W/cm$^2$
for most of our calculations. %
Additionally, we applied also pulses with $S_{\rm peak}\approx 5\times 10^{11}\,$W/cm$^2$ and $S_{\rm peak}\approx 5\times 10^{10}\,$W/cm$^2$, 
to assess the effect of nonlinear contributions (a detailed discussion is given in the Supplemental Material \cite{cn:supp}). 
The shape of the laser pulses, as well as the Fourier transform of the time-dependent electric field are shown in Fig. \ref{E-t-fourier}. Due to their finite duration,
the pulses are not monochromatic and the Fourier transform of the time-dependent electric field $\bold{E}(t) = -\partial \bold {A_{\rm ext}(t)}/\partial t$
is characterized by a Gaussian energy distribution with a constant FWHM of 0.63\,eV centered around the energy of the  monochromatic light wave.

Our present investigation covers photon energies below ($\hbar\omega=$2.25 eV), 
in the order of ($\hbar\omega=$4.5 eV) and above ($\hbar\omega=$7.75 eV) the LDA band gap of MgO
(4.64\,eV, consistent with previous results
\cite{cn:wang2004electronic,cn:schleife2006first,cn:shishkin2007self,cn:nourozi2019electronic})
with particular emphasis on the impact of the polarization direction of the light wave relative to the orientation of the surface, as indicated by the oscillating red arrows in Fig.~\ref{fig:intro}.

\begin{figure}[!htp]
\includegraphics[width=0.45\textwidth]{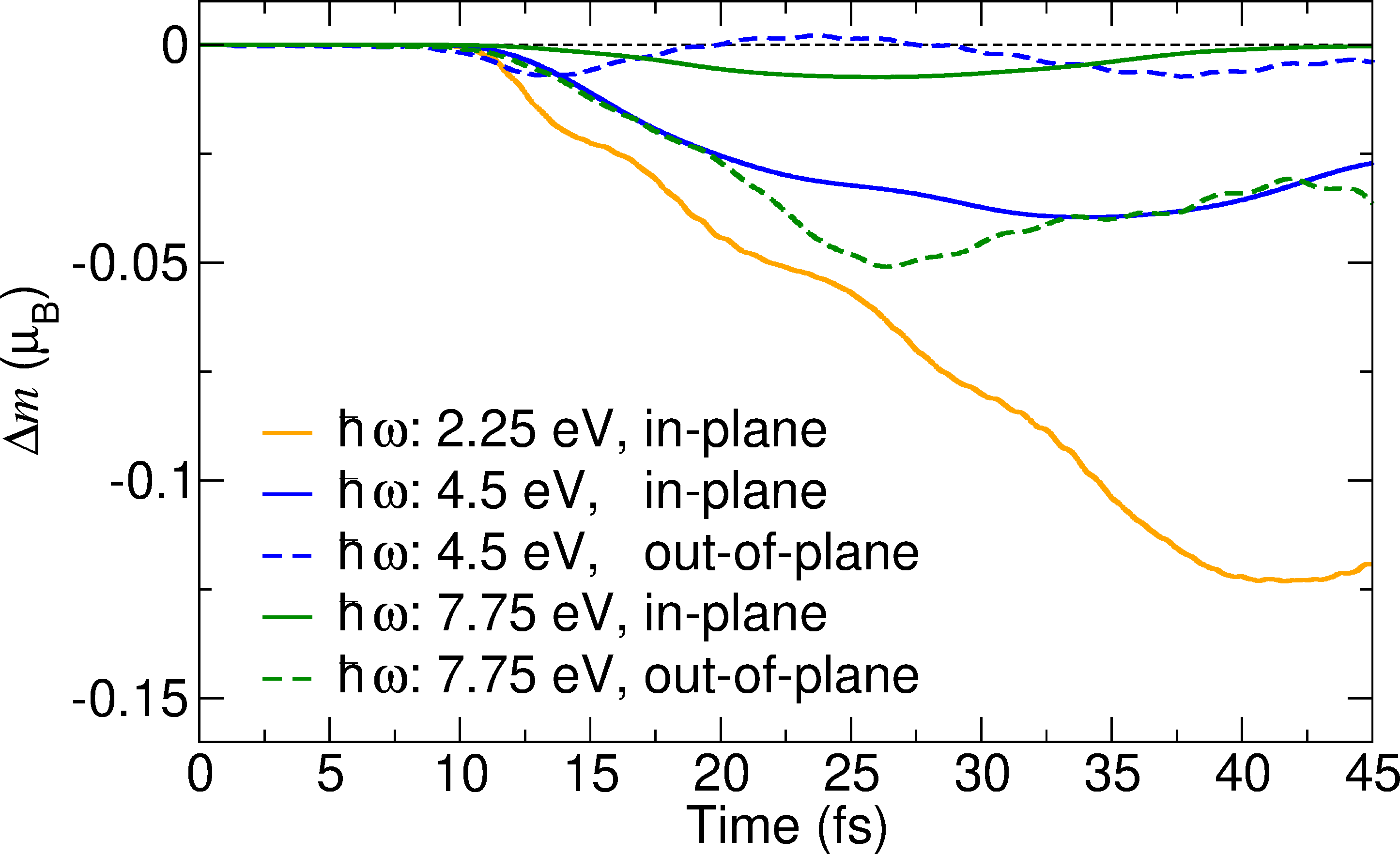}%
\caption{
  Change of the total magnetic moment in the simulation cell
  as a function of time for
in-plane and out-of-plane laser pulses with different laser frequencies.}
\label{moment}
\end{figure}
\subsection{\label{sec:mdyn} Magnetization dynamics}
Beyond our previous work  \cite{gruner2019dynamics}, we include here the spin-orbit interaction (SOI).
With SOI, the $z$-component of the electronic spin is no longer a conserved quantity, and thus the laser excitation may impact the magnetization of the system.
We expect the largest effect for Fe, which in the ground state carries a moment of
$2.25$ \mub\ (close to the experimental value for bulk bcc iron,
$2.22$ \mub\ \cite{cn:LandoltBornstein2014})
and provides nearly the complete magnetization of the system, whereas on the neighboring O- and Mg IF-sites
there is only a small induced spin-polarization of $6.4\times 10^{-4}$ \mub\ and $1.3\times 10^{-2}$ \mub, respectively. 
The time-dependent change of the total magnetization (cf.\ Fig.~\ref{moment}) shows dependence
on the laser frequency as well as polarization,
but remains with at most $\Delta m=-0.15$ \mub\ small compared to the total moment in the cell.
For an in-plane orientation of the electric field vector, the largest decrease is observed for a photon energy of $2.25\,$eV with its maximum change
around $40\,$fs after the start of the pulse and a steady recovery afterwards.
With increasing photon energies the effect is even smaller and the maximum change is reached earlier in time. This is attributed to the fact that the $d$-electrons, responsible for the large spin polarization, are more efficiently addressed by the lower energy pulses.
For a polarization of the electric field along the stacking direction of the layers,
the response of the magnetic subsystems is weaker with minute changes in magnetization for the photon energy of $4.5\,$eV and somewhat larger for $7.75\,$eV ($\Delta m=-0.05\,$\mub{}).
Thus, we conclude that SOI has only a minor impact on the dynamics of charge transfer and
excitations across the interface in \femgo. This is further supported by a detailed comparison
of the time- and layer-resolved DOS for two relevant cases in the Supplemental Material \cite{cn:supp},
which does not exhibit a notable difference between the excitation patterns obtained with SOI and within the scalar-relativistic approach.
The limited impact of the laser pulse on the magnetization dynamics indicates
that the mixing between the spin channels is small and thus
the $z$ component of the spin is rather conserved.
This allows us to use the differences between the spin-up and down projected time-resolved DOS to
identify the propagation of excitations from the strongly spin-polarized Fe layer
to the non-spin-polarized MgO part.
However, further studies involving thicker heterostructures may be necessary to address other relevant mechanisms that can lead to laser induced magnetization reversal such as the optically induced spin transfer (OISTR) mechanism \cite{dewhurst2018laser}, observed in combined ferro-/antiferromagnetic heterostructures.


\subsection{\label{sec:inplane} Laser excitations below and above the MgO gap}
In the following we will address how the response of a metal/insulator heterostructure
depends on the frequency of the electromagnetic wave.
We elucidate different scenarios in our RT-TDDFT simulation by applying
three different laser pulses with the same shape and peak power density
$S_{\rm peak}\approx 5\times 10^{12}$ W/cm$^2$ as in the previous section,
but with different excitation energies $\hbar\omega = 2.25\,$eV, $4.5\,$eV and $7.75\,$eV, covering the relevant range of photon energies. 
We start with the electric field of the pulse oriented along the $x$-axis, i.\,e., parallel to the stacking plane.
For the first energy, $\hbar\omega$$\,=\,$$2.25\,$eV, the imaginary part of the dielectric tensor exhibits a local maximum.
This energy is clearly below the band gap and can only address interface states in the MgO part or the conduction band across the charge transfer gap.

The second energy of $4.5\,$eV falls slightly short of 
the LDA band gap of $4.64\,$eV. Although the finite pulse width leads to a 
Gaussian distribution of energies with a tail above this value (cf.\ Fig.~\ref{E-t-fourier}),
direct excitations across the gap are negligible.
The largest energy, $\hbar\omega$$\,=\,$$7.75\,$eV, is on the other hand
sufficient for optical excitations in bulk MgO
(see Fig.\ S3 in the Supplemental Material \cite{cn:supp} for a comparison
of the response of bulk MgO to pulses with $\hbar\omega=4.5\,$eV and $7.75\,$eV).

\begin{figure}
\includegraphics[width=0.5\textwidth]{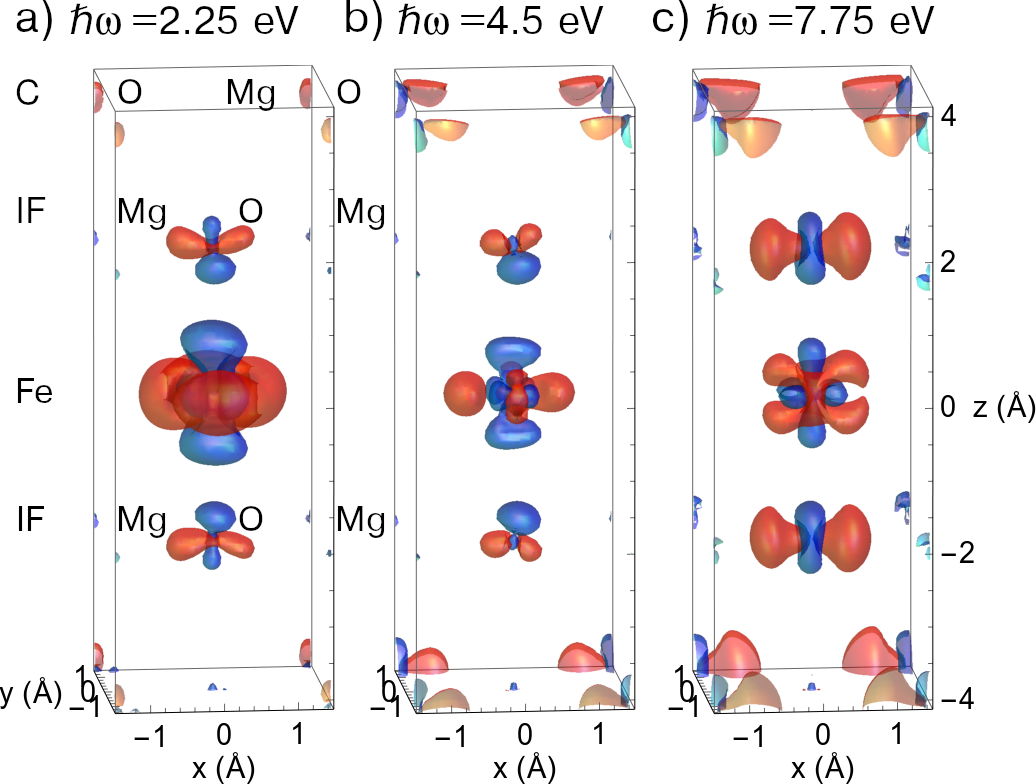}%
\caption{Changes in the charge distribution $\Delta\rho({\mathbf r},t)=\rho({\mathbf r},t)-\rho({\mathbf r},0)$ at $t=20.2\,$fs
  after illumination with an in-plane polarized laser pulse with
  $S_{\rm peak}\approx 5\times 10^{12}\,$W/cm$^2$
  for a) $\hbar\omega$ = 2.25 eV, b) $\hbar\omega$ = 4.5 eV and c) $\hbar\omega$ = 7.75 eV. Red/blue isosurfaces indicate regions with
a depletion/accumulation of charge with
  an isosurface level of $\pm 2\times10^{-3} e_0/a^3_B$.
}  
\label{fig:ChgIP}
\end{figure}

The transient charge redistribution in the Fe/MgO heterostructure,
$\Delta\rho({\mathbf r},t)=\rho({\mathbf r},t)-\rho({\mathbf r},0)$, shown in
Fig.~\ref{fig:ChgIP} for the three excitation energies at $t=20.2\,$fs, right after the decay of the laser pulse,
provides a spatially resolved illustration of
the characteristic differences between the three excitation frequencies.
For all three frequencies, only small features show up at the Mg sites,
indicating the primary importance of Fe and O orbitals.
The changes in the charge cloud around the central Fe atom are largest for $\hbar\omega = 2.25\,$eV,
while for $\hbar\omega = 7.75\,$eV we see enhanced excitations in the MgO layers,
in accordance with our considerations from the last paragraph.
The isosurfaces for $\hbar\omega$$\,=\,$$2.25\,$eV and $4.5\,$eV
indicate a transfer of charge mainly from in-plane (\dxtyt\ and/or \dxy) to \dztrt\ orbitals of Fe. Likewise at the apical O(IF) a transfer from in-plane oriented $p_x$ to out-of-plane $p_z$ orbitals takes place.
For $\hbar\omega$$\,=\,$$7.75\,$eV the pattern in $\Delta\rho({\mathbf r},t)$
rather suggests a transfer from \dxz\ and \dyz\ to \dztrt\ and in-plane $d$-orbitals at the Fe site, whereas at O(IF) and O(C)  a rather symmetric and much larger pattern emerges, indicating again a significant transfer from in-plane to out-of-plane oriented $p$-orbitals. 

\begin{figure*}
\includegraphics[width=0.9\textwidth]{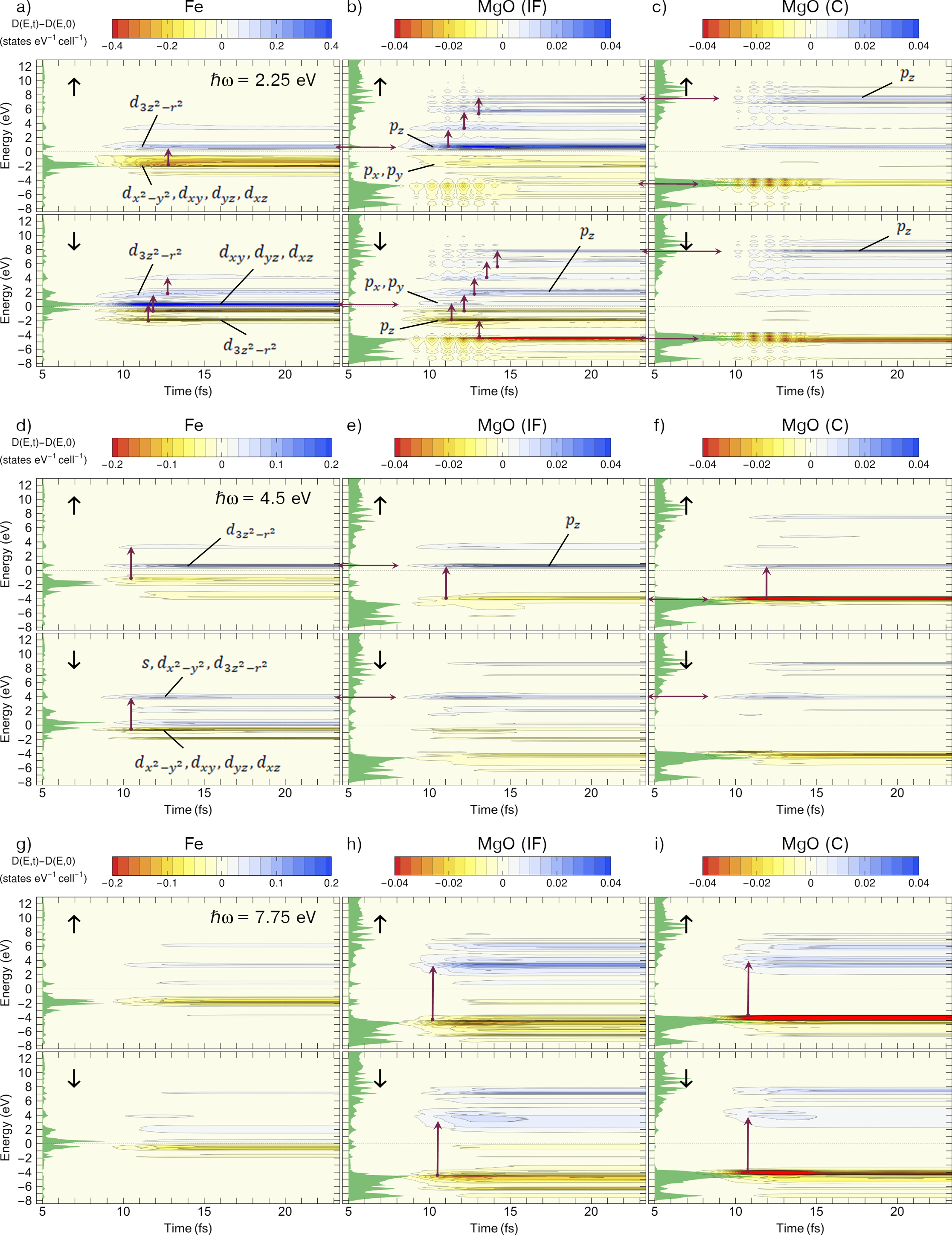}%
\caption{
  Time-dependent changes in the layer-resolved $\Delta$TDDOS
for three in-plane laser pulses with frequencies $\hbar\omega$ = 2.25 eV (a-c), $\hbar\omega$ = 4.5 eV (d-f), $\hbar\omega$ = 7.75 eV (g-i) and a peak power density of $S_{\rm peak}\approx 5\times 10^{12}$ W/cm$^2$. The three columns show the partial $\Delta D_{\sigma}(E,t)$ of the Fe, the IF-MgO and  the Center-MgO layers, respectively. The energy $E$ is given relative to the Fermi level. Note that the color scale differs between the columns. Purple arrows mark particular transitions and hybritizations which are discussed in the text.  The green area at the left edge of each panel shows the static ground-state partial density of states for the respective layer. For better visibility, the static DOS of the  MgO layers in the center and right column is scaled by a factor 4 compared to Fe in the left column.}
\label{orbital-deltaTDDFT-atom-resolved-in-plane}
\end{figure*}
To address  the question whether the excitations  are coupled or occur independently in each layer
we compare in Fig.~\ref{orbital-deltaTDDFT-atom-resolved-in-plane} the change in the spin- and layer-resolved projected time-dependent DOS $\Delta D_{\sigma}(E,t) = D_{\sigma}(E,t)-D_{\sigma}(E,0)$ for the three cases.
For the lowest frequency $\hbar\omega=2.25\,$eV, we consistently observe the largest changes in the minority spin channel of the Fe-layer, as it offers in comparison to the majority channel
a larger number of potential initial and final states.
Most of the features can be traced back to direct resonant transitions from occupied to unoccupied states in the static DOS,
but several features appear also at energies significantly above and below the Fermi level.
Such excitations are in particular present in both MgO layers, for instance close to the valence band edge of MgO, which lies around $4\,$eV below the Fermi level in the present heterostructure,
suggesting a multi-photon excitation process.
These excitation patterns disappear, when the intensity of the pulse -- specified here in terms of its peak power density -- is reduced by one or two orders of magnitude, while the features within $\pm\hbar\omega$ from \ef{} are still present (see Fig.~S6 and S7 in the Supplemental Material \cite{cn:supp} for a detailed discussion). This proves that the features far away from \ef{} arise from non-linear processes,
while resonant excitations in the vicinity of \ef{} remain the main channel
for carrier excitation.
This indicates in turn that non-linear excitation effects in sufficiently strong pulses may
be employed to  effectively
(de-)populate spatially extended states  in the valence or conduction band of MgO(IF) and MgO(C) at energies beyond the reach of a direct excitation.
Thus, for MgO(C) a significant depletion of valence band states takes place between $-4$ and $-5\,$eV, almost as large as in MgO(IF). Likewise, we observe connected features at $+8\,$eV, both
indicated by the horizontal arrows in Fig.~\ref{orbital-deltaTDDFT-atom-resolved-in-plane}(b-c).
In MgO(C) intermediate levels, from which carriers might by
excited beyond the gap, are essentially absent owing to the exponential decay of interface states in the band gap with increasing distance from the interface. Nevertheless, the changes in occupation numbers in the valence and conduction band are even larger in the central layer compared to the interface.

The resonant excitations in the vicinity of the Fermi level confirm our previous conjecture, that hybridization between out-of-plane oriented orbitals in adjacent layers is particularly relevant for the transfer of excitations into and through the interface \cite{gruner2019dynamics}, see
Fig.~\ref{orbital-deltaTDDFT-atom-resolved-in-plane}(a-c).
Because of the large exchange-splitting, unoccupied
\dztrt\ states of Fe in the majority spin channel are located in the right window,
$+0.8\,$eV above \ef{},
such that electrons can be transferred from in-plane
$d$-orbitals around $-1.5\,$eV below \ef\ (vertical arrows in Fig.~\ref{orbital-deltaTDDFT-atom-resolved-in-plane}(a)) to
\dztrt. These hybridize with the $p_z$ orbitals of MgO(IF), which show a significant occupation as well
in this energy range.
In contrast, the decrease in occupation of the $p_x$ and $p_y$ states in MgO(IF) around $-1.5\,$eV
appears much weaker, in particular compared to the rather strong effect in
the \dxtyt\ and \dxy\ orbitals of the Fe layer found at the same energy.
In the minority channel, the out-of-plane oriented bonding \dztrt\ states at $-2\,$eV
mediate the excitation of carriers from the O $p_z$ states in the interface
to the Fe \dxz\ and \dyz\ states around $+0.5\,$eV, which
hybridize with the in-plane oriented $p_x$ and $p_y$ interface states in MgO(IF),
see the horizontal arrows in Fig.~\ref{orbital-deltaTDDFT-atom-resolved-in-plane}(a-b).
Their occupation is much lower, compared to the out-of-plane
$p_z$ states in the majority channel in the same energy range.


The excitation pattern becomes more defined when we increase the energy towards the band gap of MgO.
Due to the confined width of the $d$ band in the Fe-monolayer, the number of matching initial and final states for excitations within the $d$-band of Fe is diminished.
For $\hbar\omega=4.5\,$eV, our simulations do not yield significant direct excitations between valence and conduction band states in bulk MgO.
Thus, at this energy we can pinpoint the interface layer as pivotal for the excitations.
Accordingly, we still observe in all layers a different excitation pattern in the minority and majority channel arising from the proximity to the spin-polarized Fe-layer.
In the majority channel, carriers from the delocalized valence band edge
of MgO(IF) and MgO(C) are excited into unoccupied interface states at $+0.5\,$eV, which hybridize with the \dztrt\ states in the Fe-layer.
In the minority channel, we observe a similar mechanism as described above, which effectively relocates charge density in reverse direction, towards MgO. It involves the 
\dztrt\ states of Fe which hybridize with conduction band states of
MgO(IF) and MgO(C), see Fig.~\ref{orbital-deltaTDDFT-atom-resolved-in-plane}(d-e).

If we take a separate look at the central layer
in Fig.~\ref{orbital-deltaTDDFT-atom-resolved-in-plane}(f), we observe that
states at the valence band edge of MgO(C) deplete and a concomitant occupation in the conduction band takes place, which is, however, separated by at least twice the laser energy. The comparison with our calculations for MgO bulk prove (cf.\ Fig.\ S3(b) in the Supplemental Material \cite{cn:supp}), that the intensity of the
excitation cannot be explained by a direct transition in bulk MgO.
Instead, the IF layer plays a decisive role, as it mediates concerted
excitations from the valence band states extending between MgO(IF) and MgO(C) to the hybridized states of MgO(IF) and Fe just above the Fermi level and, simultaneously, hybridized states of MgO(IF) and Fe just below the Fermi level to extended conduction band states in MgO(IF) and MgO(C). These processes work in opposite direction and can be invoked independently by the same laser pulse. Their combination prevents an effective charge transfer between the metallic
and the insulating subsystem.

For $\hbar\omega=7.75\,$eV, a substantial laser excitation can occur directly in the MgO subsystem without the support of interface states.
As a consequence, the excitation patterns of both spin channels assimilate in both layers of MgO, as compared to the lower photon energies.
The largest fraction of the excitations removes states from the first peak below the valence band edge at $-4\,$eV (corresponding to $-1.2\,$eV in bulk MgO) and populates states around $3.5$-$4.0\,$eV (corresponding states in bulk MgO are located around $6.5\,$eV, see Fig.\ S3(a) in the Supplemental Material \cite{cn:supp}), which are marked by the vertical arrows in Fig.~\ref{orbital-deltaTDDFT-atom-resolved-in-plane}(g-i).
Direct excitations in the Fe-layer, as observed for the smaller frequencies, are largely absent. The substantial changes in the occupation numbers in Fe close to the Fermi level are again best explained through
the interface-mediated mechanisms discussed above, which originate from the
the hybridization between the spin-polarized states in Fe with MgO(IF).
Despite the largest amount of excitation taking place directly in MgO(IF) and MgO(C), there is still a notable remaining asymmetry between the spin channels visible in Fig.~\ref{orbital-deltaTDDFT-atom-resolved-in-plane}(i). This asymmetry
is a consequence of the proximity to the spin-polarized metal layer,
since the corresponding static DOS of both spin channels is rather similar
(cf.\ the green areas in Fig.~\ref{orbital-deltaTDDFT-atom-resolved-in-plane}).

\subsection{\label{sec:outplane} Polarization dependence of the excitation pattern}

\begin{figure}[!htp]
\includegraphics[width=0.5\textwidth]{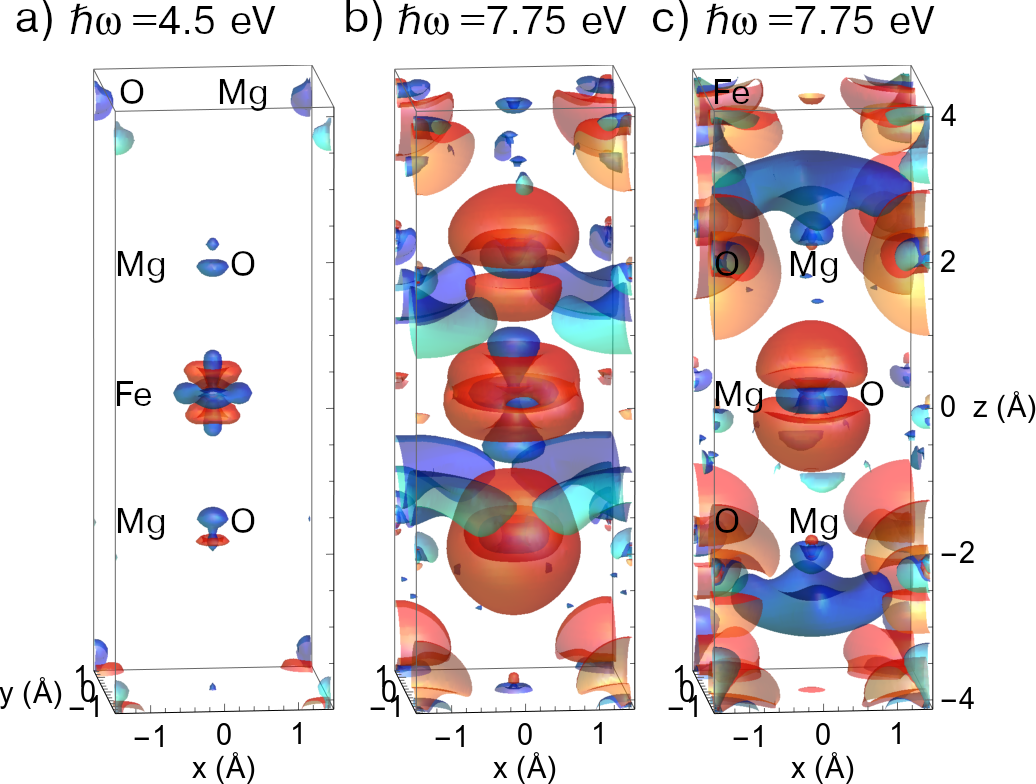}%
\caption{
  Snapshot of the evolution of the charge distribution
  $\Delta\rho({\bold r},t)$
  for out-of-plane polarized pulses at $t=20.2\,$fs for
  $\hbar\omega = 4.5\,$eV in (a) and $\hbar\omega = 7.75\,$eV in (b). Same colors and
  isolevels as in Fig.~\ref{fig:ChgIP}.
  The positions of the atoms in the cell in (a) and (b) are the same as in Fig.~\ref{fig:intro} with Fe in the center. In (c) we shift the differential charge distribution shown in (b) by 0.5 lattice vector in each direction, in order to visualize better the features at the edges of the original unit cell. As a result, O(C) is now located in the center, while Fe is placed at the edges of the cell.}
\label{z-Q}
\end{figure}
\begin{figure*}
\includegraphics[width=1\textwidth]{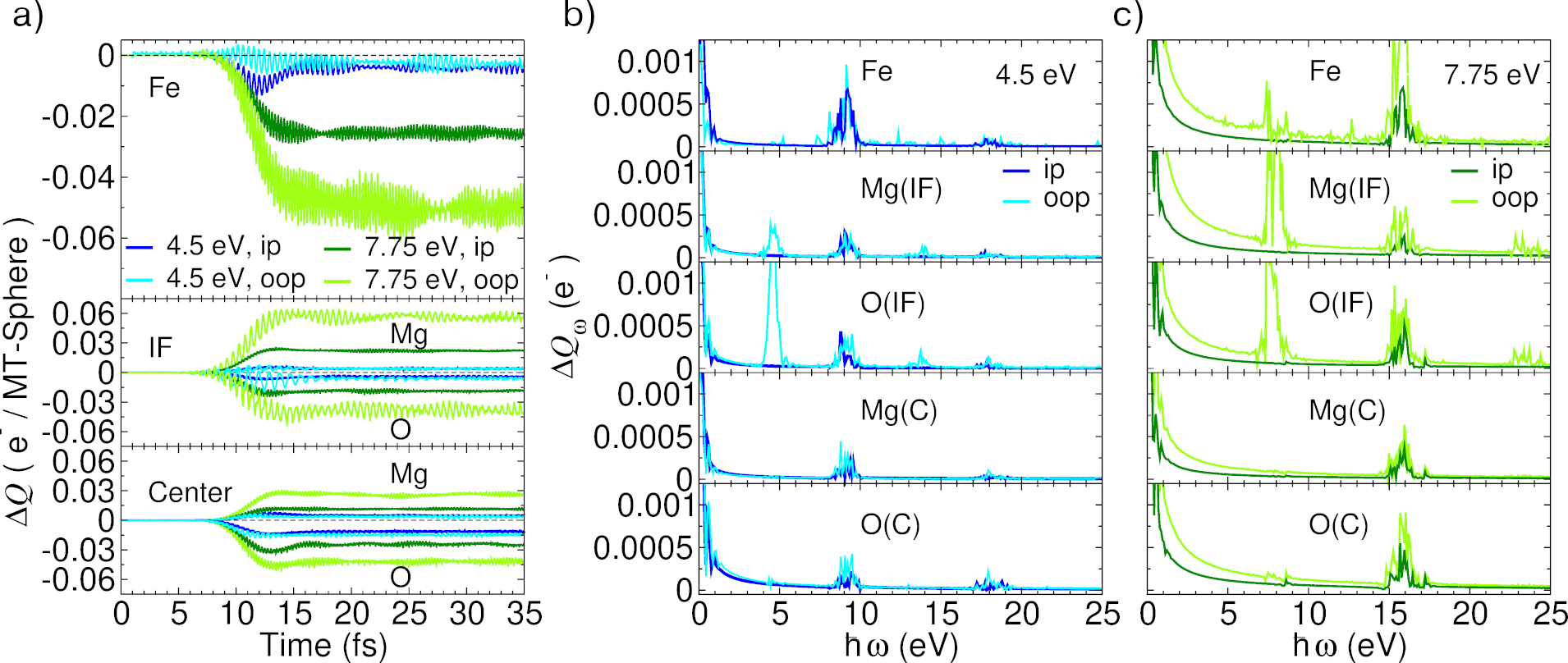}%
\caption{
  a) Change of the electronic charge inside the muffin-tin spheres around the ions relative to the initial state, $\Delta Q(t) = Q(t)-Q(0)$  during and after the application
  of in-plane (ip) and out-of-plane (oop) polarized
  laser pulses with $\hbar\omega$ = 4.5 eV and $\hbar\omega$ = 7.75 eV.
  The initial charges $Q(0)$ within the muffin tin spheres for Fe, Mg(IF), O(IF), Mg(C) and O(C) are
  $24.42\,e_0$, $10.67\,e_0$, $7.24\,e_0$, $10.72\,e_0$ and $7.24\,e_0$, respectively.
  b) and c) 
Fourier transform of $\Delta Q(t)$ shown in a) for $\hbar\omega$ = 4.5 eV and $\hbar\omega$ = 7.75 eV, respectively.}
\label{Q-fourier}
\end{figure*}

The diagonal components of the imaginary part of the dielectric tensor Im[$\epsilon(\omega)$]
reported previously \cite{gruner2019dynamics}
indicate a predominance of  in-plane components for low excitation energies and  
a substantial increase of absorption in $z$-direction for frequencies above the band gap of MgO.
In particular we found that for our system, the 
in-plane component Im[$\epsilon_{xx}(\omega)$] resembles bulk Fe, while the out-of-plane component
Im[$\epsilon_{zz}(\omega)$] rather bears similarity to bulk MgO, in particular regarding the
low absorption at energies below the gap. 
This implies a strong dependence of the excitation pattern on the polarization of the electric field component of the light pulse.

To assess this, we carried out additional RT-TDDFT simulations for $\hbar\omega=4.5\,$eV and $7.75\,$eV
with the vector potential (and thus the electric field) 
pointing along the $z$-axis, \ie\ along the stacking sequence of the heterostructure.
The peak power density $S_{\rm peak}\approx 5\times 10^{12}\,$W/cm$^2$ and pulse shape was kept the same as in the last section.
The corresponding differential densities taken at 20.2\,fs are shown in Fig.~\ref{z-Q}. They exhibit indeed
substantial differences to the in-plane polarized field but also between the two laser frequencies. At $4.5\,$eV we observe a much weaker response compared to the in-plane case, which involves mainly a redistribution of charge from \dxz\ and \dyz\ to \dztrt\ and \dxtyt\ orbitals at Fe, cf.\ Fig.~\ref{z-Q}(a), similar to what was found for $\hbar\omega=1.67\,$eV in Ref.\ \onlinecite{gruner2019dynamics}.
In contrast, Fig.~\ref{z-Q}(b,c) shows a substantially increased charge redistribution at all sites in the unit cell for the out-of-plane polarized 7.75\,eV pulse as compared to its in-plane counterpart (cf.\ Fig.~\ref{fig:ChgIP}(c)).
At the Fe-site we monitor once again an occupation of \dztrt\ states and a depletion of orbitals with in-plane-character, whereas the pattern at the apical and central O-sites suggest a redistribution between $s$ and $p_z$ orbitals. In contrast to frequencies below the bulk band gap of MgO, we encounter rather significant changes in the charge distribution around Mg(IF) sites as well. These effects are rather asymmetric and extend towards the Fe site into the interstitial.

The characteristic features of the  differential charge density $\Delta\rho$ are also
reflected in the integrated charge density  $\Delta Q$ in the muffin tin spheres around each site,
shown in Fig.~\ref{Q-fourier}(a).
We encounter for all frequencies and both directions of polarization a steep increase in $\Delta Q$ at
the onset of the pulse, followed by a decrease, reaching a steady value after the pulse.
At the O sites, we see a corresponding decrease in  $\Delta Q$, accompanied by
an increase of similar magnitude at the Mg sites, which we previously interpreted as an effective charge transfer from O to Mg  \cite{gruner2019dynamics}.
Consistently, this charge transfer increases significantly
for $\hbar\omega=7.75\,$eV where direct excitations within bulk MgO occur.
For the out-of-plane polarized pulse the increase of charge around Mg(IF) is significantly larger than
the decrease around O(IF). Simultaneously,  the amount of charge leaving the muffin tin sphere of Fe has substantially increased. This suggests that the large blue features close to Mg(IF)
in Fig.~\ref{z-Q}c arise from a relocation of charge
from the 
$d$ orbitals of Fe.

The $\Delta Q$ are modulated by an oscillation with twice the frequency of the laser excitation,
alluding a coherent ringing of the charge clouds which extends
until the end of our simulations at 45\,fs. This is combined with characteristic beats indicating
the superposition of oscillations with slightly varying frequencies.
The frequency doubling is a consequence of the electric field driving the charge out and back into the muffin tin sphere at both sides in a similar way
at positive and negative amplitudes. This symmetry is, however,
broken at the interface for out-of-plane oriented pulses and the oscillations for Mg(IF) and O(IF)
exhibit only half the frequency of the Fe and the central layers of MgO, accordingly.
This is substantiated by the Fourier transform $\Delta Q(\omega)$ shown in Fig.~\ref{Q-fourier}(b) and 
\ref{Q-fourier}(c) for 
$\hbar\omega=4.5\,$eV and $\hbar\omega=7.75\,$eV, respectively. All curves show a large contribution at
$\omega=0$ and features at $2\hbar\omega$ with a finite extension, similar to the
width of the laser pulse. The latter can explain the beating in the oscillations in $\Delta Q(t)$. For 
$\hbar\omega=4.5\,$eV we also observe higher harmonics at $4\hbar\omega$, which appear most pronounced
for O(C). In addition, the out-of-plane polarized pulses shown in Fig.~\ref{Q-fourier}(b) and \ref{Q-fourier}(c) 
exhibit large peaks 
at $\hbar\omega$ with a satellite at $3\hbar\omega$ in particular in the interface MgO layer. These peaks are suppressed in Fe and the central layer of MgO, which both have mirror symmetry.

\begin{figure*}
\includegraphics[width=1\textwidth]{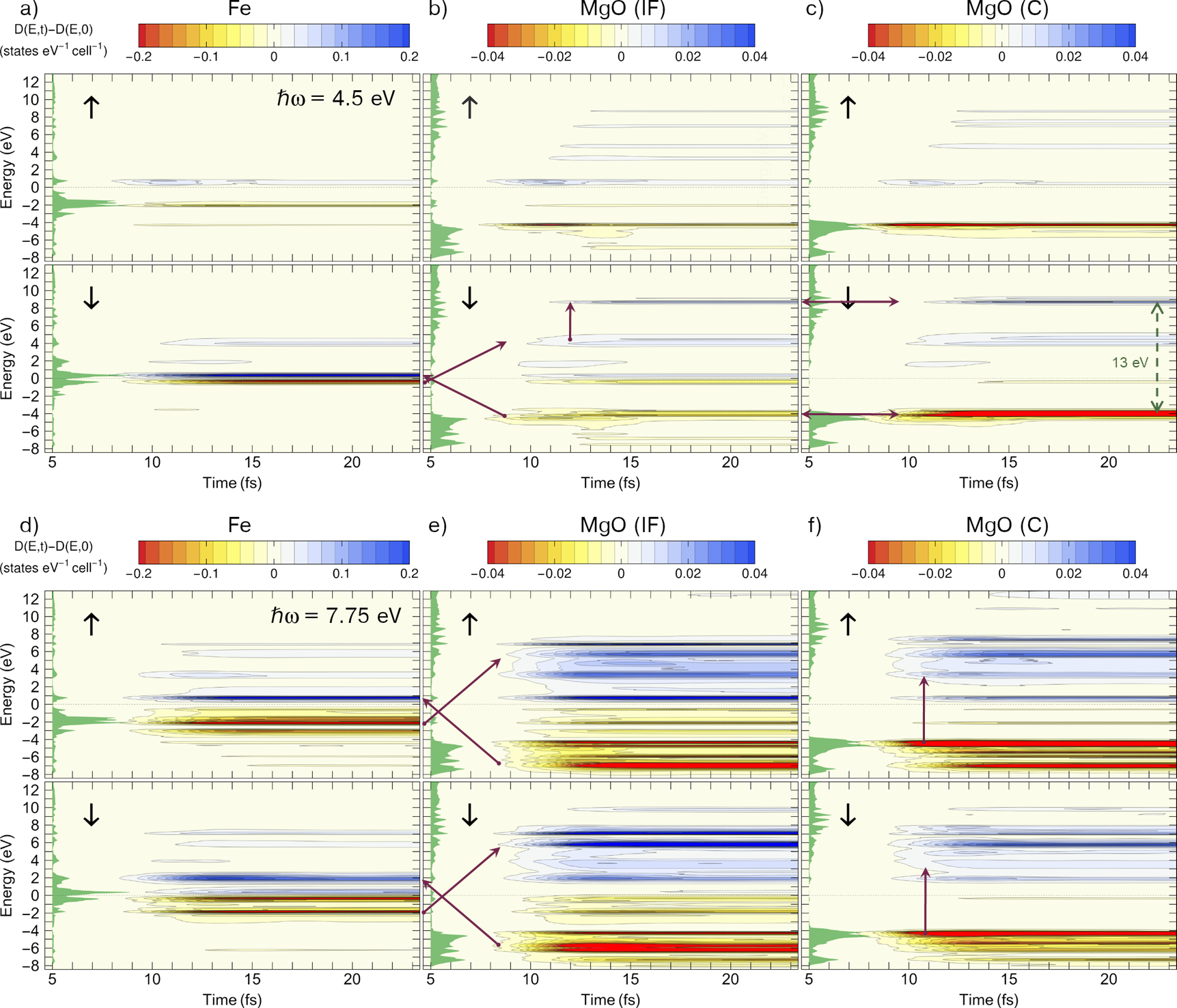}%
\caption{
  Time-dependent changes in the layer-resolved $\Delta$TDDOS
for out-of-plane laser pulses with frequencies $\hbar\omega = 4.5\,$eV (a-c)
and $\hbar\omega= 7.75\,$eV (d-f) and a peak power density of
$S_{\rm peak}\approx 5\times 10^{12}$ W/cm$^2$. The three columns show the partial $\Delta D_{\sigma}(E,t)$ for the Fe, MgO(IF) and the MgO (C) layers, respectively.
The energy $E$ is given relative the Fermi level.
Note that the color scale differs between the columns.
Purple arrows mark particular transitions and hybritizations which are discussed in the text.
The green areas refer again to the static partial DOS, scaled as in Fig.~\protect\ref{orbital-deltaTDDFT-atom-resolved-in-plane}.
}
\label{orbital-deltaTDDFT-atom-resolved}
\end{figure*}

The particular role of the interface for the excitation is even more pronounced for the out-of-plane polarization of the electric field,  as shown in Fig.~\ref{orbital-deltaTDDFT-atom-resolved}.
Most conclusive is once again the minority channel, where occupied and unoccupied $d$ states of Fe
are nearly balanced.
For $\hbar\omega=4.5\,$eV, we find here two pronounced and rather sharp features above and below the Fermi level, which
are too close together to be explained by resonant excitations within the Fe layers,
see Fig.~\ref{orbital-deltaTDDFT-atom-resolved}(a).
However, taking into account that these features hybridize with interface states in MgO(IF),
we can conjecture once again two transitions taking place cooperatively,
which result in a charge transfer into the empty $3d$ orbitals of Fe:
One from the valence band of MgO into the interface states above \ef\ and another one from the interface states below \ef\ into the conduction band of MgO, see Fig.~\ref{orbital-deltaTDDFT-atom-resolved}(a-b).
The latter is involved in a subsequent non-linear excitation process, populating the MgO levels at 8.5\,eV.
Furthermore, we observe similar features with larger changes in the occupation numbers above and below the Fermi level for the central layer of MgO, see Fig.~\ref{orbital-deltaTDDFT-atom-resolved}(c),
keeping in mind that 4.5\,eV is not yet sufficient for a substantial
direct excitation across the band gap.
Interestingly, in MgO(C) the population of the unoccupied states at $8.5\,$eV appears to be even more defined compared to the pattern at $4\,$eV, which we can link to a direct excitation in the IF layer. This indicates, that excitations in the non-linear regime may actively foster the transfer of excitations across the IF.

As expected from the differential charge density and the larger absorption coefficient for out-of-plane light, $\hbar\omega=7.75\,$eV leads in all three distinct layers to a much richer pattern in $\Delta D(E,t)$.
In Fig.~\ref{orbital-deltaTDDFT-atom-resolved}(f), which refers to MgO(C), the vertical arrows denote the corresponding transition from the valence to the conduction band in bulk MgO.
Excitations
above $+6\,$eV and below $-6\,$eV apparently involve transitions to and from the interface states of MgO(IF).
Consequently, we see changes in the occupation numbers of MgO(IF) and Fe,
which are consistent with the mechanism identified above:
From approximately $-6\,$eV to $+1\ldots 2\,$eV and from $-2\,$eV and above to $+6\ldots 7\,$eV, as indicated by the diagonal
arrows in Fig.~\ref{orbital-deltaTDDFT-atom-resolved}(d-e):
While transitions to the highest and from the lowest energies are exclusive to MgO(IF) and MgO(C)
excitations in the vicinity of the Fermi energy accumulate again in the hybridizing $d$ states of Fe, which provide a large density of states and can serve as a buffer for the excitations from and into the MgO subsystem.

The combination of a $d$-metal and a wide-band-gap insulator leads thus not only to a transfer of excited carriers into the subsystems
but also to an asymmetric transfer of energy. In the Fe layer, due to the confinement, excited electron and hole states lie closer together
than the photon energy. The distribution of excitations
is thus closer to thermalization than one would obtain from an excitation of bulk Fe
with the same light pulse. In contrast, excited positive and negative carriers are separated by up to $13\,$eV in MgO(C) for a strong $4.5\,$eV light pulse.

\section{\label{sec:sum} Discussion and Conclusion}
\begin{figure}
\includegraphics[width=\columnwidth]{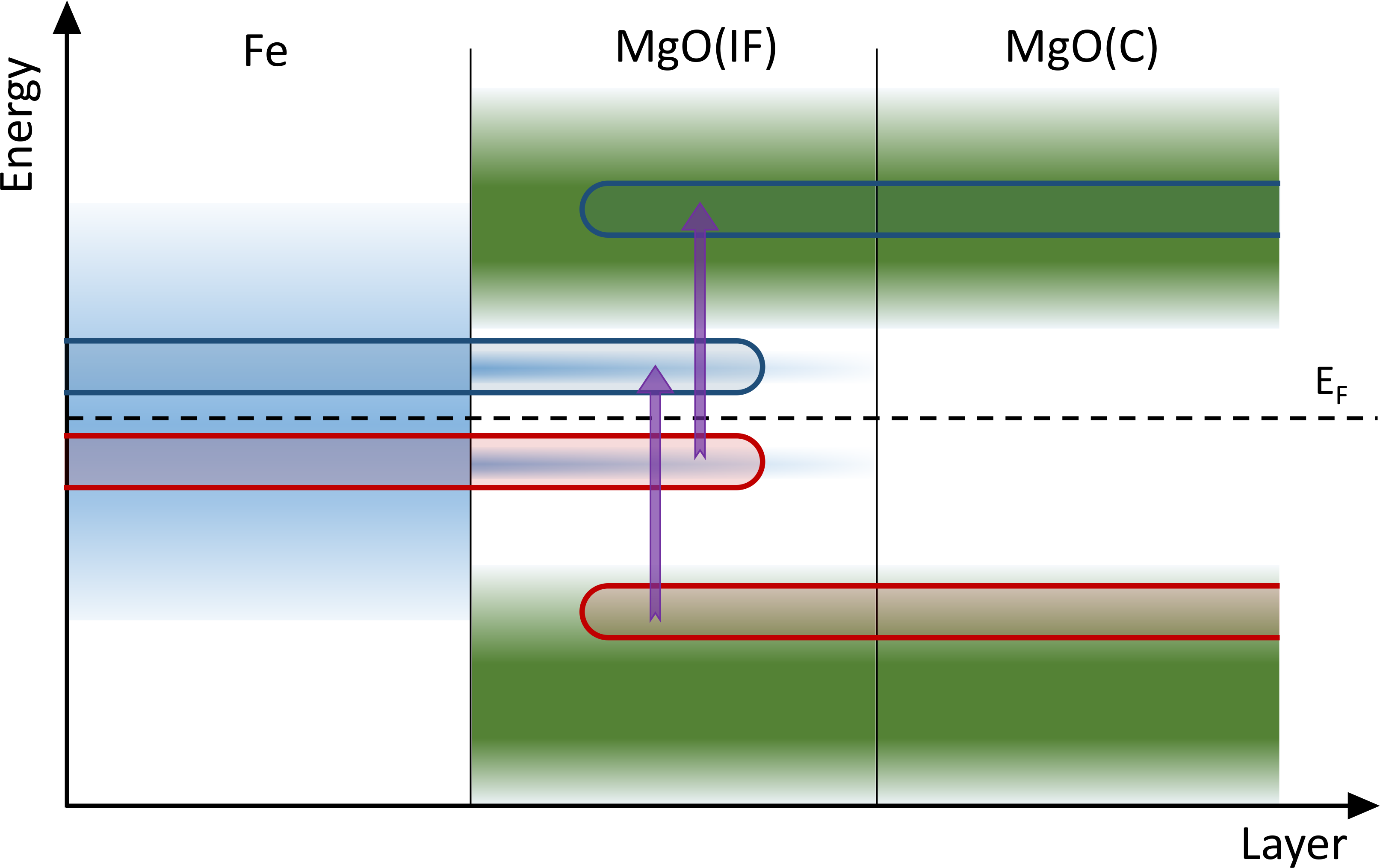}%
\caption{
  Schematic illustration of the concerted excitation mechanism
  mediated by interface states involving
  two simultaneous, but independent excitations across the charge transfer gap.
  This leads to the accumulation of
  excited carriers with a small energy separation in the metallic layer and
  a significantly larger separation than $\hbar\omega$ in the MgO part. In this way an efficient
  repopulation of carriers from valence band (red box) to conduction band states
  (blue box) is possible even
  for photon energies below the band gap of the insulator.
  The colored areas refer to the electronic bands. Green: valence and conduction band of the (bulk) insulator; blue: $d$-states of the (bulk) metal and interface states. 
}
\label{fig:mechanism}
\end{figure}

Our systematic RT-TDDFT investigation of a
minimal model Fe/MgO heterostructure shows that the response of the system
strongly depends on both the photon frequency and polarization of the light field.
Moreover, we could unravel the origin of the excitation and the transfer of
carriers in full detail.
In contrast to our previous study on \femgo, we considered spin-orbit interactions and found that demagnetization effects do not exceed 10\,\% of the ground state magnetization of the Fe-layer and can thus be regarded of minor
relevance in this particular system.

We find in accordance with our previous study that for photon energies substantially below the LDA band gap of MgO
excitations take place predominantly in the Fe subsystem, which provides a rich variety of initial and final states within the
width of the $d$-band. This works most effectively when
the electric field component of the light lies within the plane of the Fe-layer.
The situation changes drastically when the photon energy approaches
or even exceeds the LDA band gap of MgO. The reduced dimension
of the Fe-layer narrows the d-band width of the Fe monolayer compared to bulk, with the consequence that
even photon energies around $4.5\,$eV (i.\,e., close to the LDA band gap of MgO)
can hardly trigger substantial excitations within the Fe-layer.
Thus, for frequencies in the order of the $d$-band width and larger, the hybridization of Fe-$d$ and O-$p$
states at the interface plays a decisive role for the excitation.

Hence, we could identify a generic mechanism associated with these states, which
allows to pump excitations from deep within the valence
band into the conduction band of the insulator even when the photon
energy is not sufficient to reach the final states.
This mechanism involves two simultaneous (but not necessarily coherent)
excitations in the interface region, see Fig.~\ref{fig:mechanism}.
The first excites valence band electrons of the insulator into an interface
state above the Fermi level, while the second promotes electrons from the interface states below the Fermi level into the conduction band of MgO. The interface states of oxygen usually exhibit a low DOS, but since they hybridize with the $d$-states of the transition metal, the latter can act as a reservoir to accept and donate excited carriers to the apical oxygen.
In turn, the relevant states in valence and conduction band
can extend into the bulk layers of MgO, where direct transitions are inhibited by the band gap.
Although this takes place independently, hot electrons and holes obtained from the two processes
might thus be observed simultaneously in the bulk of the insulator,
avoiding the penalty related to the Coulomb interaction between
charged zones. Indeed, we found clear indications of such a joint
accumulation of excited holes and electrons in the central layer.
Whether a likewise transfer of excitations can be observed from the interface layer of Fe into the
bulk needs to be assessed in future studies for thicker Fe films. Here, again, a sufficient
hybridization of the out-of-plane $d$-orbitals between the inner Fe-layers and the IF-states might be
a decisive factor.
The above described mechanism leads to an asymmetric distribution of excitations, where -- directly after the pulse --
the excited electrons and holes in the metal are much closer to the Fermi level, as compared to the MgO subsystem.
This might then lead to a substantially different dynamics of thermalization
in both subsystems of the heterostructure as compared to the
respective bulk systems, 
which might be detectable in time-resolved pump-probe experiments on the fs-time scale.
Sufficiently high laser intensities, as applied in our computational approach, can furthermore trigger
a non-linear multi-photon excitation process that allows to reach
hybridized valence and conduction band states
with a significantly larger distance to the Fermi level as compared to the photon energy.
This implies that a similar effect might be achieved by the simultaneous application of two weaker
laser pulses with different, appropriately chosen frequencies.

Our simulations indicate a strong dependence of the absorbed light on the polarization of the electrical field and the frequency.
As a consequence, the polarization direction emerges as an efficient means to
select transitions to specific interface states.
For frequencies below the band gap, in-plane polarized light induces a much stronger absorption, as shown previously \cite{gruner2019dynamics},
whereas for photon energies above the band gap, the out-of-plane direction leads to a significantly larger response.
While in the former case excitations are confined to the Fe-layer, the direct excitations within the MgO play the most important role in the latter, since the finite width of the Fe-$d$ band limits the number of possible transitions in the ultrathin Fe-layer.
For photon energies in the order of the band gap, we found excitations of similar strength for both polarization directions. Still we can distinguish these cases by a different excitation pattern in Fe and the adjacent MgO-layer.
We link this to the local symmetry breaking at the interface,
leading to an asymmetric deformation of the atomic charge cloud, which follows
the oscillation of the vector potential.

Our calculations were carried out for a minimal model heterostructure.
The experimental synthesis of this structure may not be trivial,
but we believe that the above sketched mechanisms can be transferred also to larger,
more realistic systems with nanometer-sized layers, as used in previous experiments \cite{cn:Rothenbach19}.
An important consequence of the confinement
is the significantly reduced bandwidth of the $3d$-metal monolayer.
A wider $d$ band would provide according to Fermi's golden rule
additional possibilities for direct excitations within the Fe-subsystem
for a laser pulse in the range of the calculated band gap.
On the other hand, the MgO band gap is underestimated due to the LDA exchange correlation functional.
Thus for a direct comparison with experiment, a quantitative description of the band gap and the location of the interface states is desirable. This requires the appropriate treatment of exchange and correlation beyond the (adiabatic) local density approximation in a time-dependent approach for both the ferromagnetic metal and the insulator in a computationally efficient implementation and must thus be left open for future work.

Thus, despite the above mentioned restrictions, the model used in this study captures
the essential excitation processes that may also be relevant for thicker heterostructures, when a laser pulse is applied with photon energy, which is slightly lower than the band gap of the insulator, but in the vicinity of or even larger compared to the $d$-band width of the metal.
We expect that the fundamental mechanisms of excitation dynamics demonstrated
in our proof-of-principle calculations are
applicable to a wider range of  metal-insulator heterostructures,
which are characterized by a large density of states around the Fermi level
in the metallic subsystem, and thus provide important guidelines for
future pump-probe experiments.

\bigskip

\begin{acknowledgments}
We wish to acknowledge funding by the Deutsche Forschungsgemeinschaft
(DFG, German Research Foundation) within collaborative research center
CRC1242 (project number 278162697, subproject C02) and computational
time at the Center for Computational Sciences and Simulation of the
University of Duisburg-Essen on the supercomputer magnitUDE
(DFG grants INST 20876/209-1 FUGG, INST 20876/243-1 FUGG).
\end{acknowledgments}


%

\end{document}